\documentclass[12pt,aps,prd,preprint,tightenlines,superscriptaddress,showpacs,nofootinbib]{revtex4-1}
\usepackage{graphicx}

\usepackage{amsmath} %??
\usepackage{graphicx}
   \usepackage{multirow}
\usepackage{epstopdf}
\newcommand{\PRE}[1]{{#1}} % Use if preprint style

\newcommand{\be}{\begin{equation}}
\newcommand{\ee}{\end{equation}}
\newcommand{\bea}{\begin{eqnarray}}
\newcommand{\eea}{\end{eqnarray}}

\def\tev{\, {\rm TeV}}
\def\gev{\, {\rm GeV}}
\def\mev{\, {\rm MeV}}

\newcommand{\sigmaSI}{\sigma_{\rm SI}}

\newcommand{\gsim}{\lower.7ex\hbox{$\;\stackrel{\textstyle>}{\sim}\;$}}
\newcommand{\lsim}{\lower.7ex\hbox{$\;\stackrel{\textstyle<}{\sim}\;$}}

\newcommand{\pb}{{\rm pb}}

\usepackage{enumerate}
\usepackage{amssymb,amsmath}

\begin{document}

\preprint{UH-511-1239-2014}
\preprint{CETUP2014-004}

\title{
\PRE{\vspace*{1.3in}}
\textsc{Charged mediators in dark matter scattering with nuclei and the strangeness content of nucleons: Strange Brew}
\PRE{\vspace*{0.1in}}
}

\author{Chris Kelso}
\affiliation{\mbox{Department of Physics, University of North Florida, Jacksonville, FL  32224, USA}
%\PRE{\vspace*{.1in}}
}

\author{Jason Kumar}
\affiliation{\mbox{Department of Physics and Astronomy, University of
Hawai'i, Honolulu, HI 96822, USA}
%\PRE{\vspace*{.1in}}
}

\author{Pearl Sandick}
\affiliation{\mbox{Department of Physics and Astronomy, University of Utah, Salt Lake City, UT  84112, USA}
\PRE{\vspace*{.1in}}
}

\author{Patrick Stengel \PRE{\vspace*{.1in}} }
\affiliation{\mbox{Department of Physics and Astronomy, University of
Hawai'i, Honolulu, HI 96822, USA}
%\PRE{\vspace*{.1in}}
}

%--- Abstract ----------------%
\begin{abstract}

We consider a scenario, within the framework of the MSSM, in which dark matter is
bino-like and dark matter-nucleon spin-independent scattering occurs via
the exchange of light squarks which exhibit left-right mixing.  We show that direct
detection experiments such as LUX and SuperCDMS will be sensitive to a wide class
of such models through spin-independent scattering.  Moreover, these models exhibit
properties, such as isospin violation, that are not typically observed for the MSSM
LSP if scattering occurs primarily through Higgs exchange.  The dominant nuclear physics uncertainty
is the quark content of the nucleon, particularly the strangeness content.

\end{abstract}

\maketitle

%\tableofcontents

\section{Introduction}

Within the framework of the MSSM, where the lightest neutralino is a dark matter candidate,
dark matter can scatter off nuclei via the exchange of the $Z$-boson, a Higgs boson, or a squark.
If dark matter-nuclei scattering is spin-independent, it is typically assumed that scattering arises
through Higgs exchange, with the scattering cross section depending on the higgsino fraction of the
lightest neutralino.  In this work, we investigate an alternative scenario, in which the lightest
neutralino is bino-like and exhibits spin-independent scattering against nuclei via light squark
exchange.  Our goal in this work is to study a range of models, consistent with current data and the
uncertainties in nuclear physics,
for which current or upcoming direct detection experiments could observe spin-independent scattering via
squark exchange.

Under typical model assumptions, spin-independent scattering of MSSM neutralino dark matter with nuclei is significantly simplified:  because the lightest neutralino of the MSSM is a Majorana fermion, $Z$-boson exchange yields a scattering matrix element
that is largely spin-dependent, while the spin-independent piece is velocity-suppressed and is therefore subdominant.  Squark-exchange diagrams can usually be neglected
because the spin-independent part of the matrix element is proportional to the mixing between the left-handed and right-handed
light squarks;
one typically assumes minimal flavor violation (especially in simplified frameworks such as the CMSSM, and even in more general, low-scale studies of the MSSM such as the pMSSM~\cite{Berger:2008cq}), where the
mixing between left and right light squarks is heavily suppressed.
Spin-independent scattering with nuclei is then due to Higgs exchange processes, with the scattering cross section largely
determined by the dark matter mass and its bino and higgsino fraction.  Higgs exchange results in dark matter-nucleon
scattering that is largely isospin invariant.    However, if one allows deviations from minimal flavor violation,
the mixing angle for the light squarks could be large, resulting in an observable spin-independent scattering cross section due to squark exchange.
Moreover, the relative strength of the coupling of dark matter to protons and neutrons can vary over a wide range.

In this scenario, largely bino-like dark matter can potentially be observed via spin-independent scattering at direct detection experiments.  Bino-like
dark matter has recently been reconsidered in~\cite{Buckley:2013sca,Pierce:2013rda,Fukushima:2014yia}, in which it has been argued that slepton mixing can permit $s$-wave annihilation
that depletes the dark matter density enough to obtain consistency with cosmological observations, thus reopening the ``bulk"
region of MSSM parameter space.  We will see that the parameter space of these bino-like models factorizes into several sectors; while the slepton
sector is relevant for determining the dark matter density, the light squark sector is relevant for dark matter direct detection.

A large uncertainty in the the spin-independent scattering cross section of dark matter with nuclei arises from the strange quark content of the
nucleon.  Results from chiral perturbation theory have long led to the conclusion that strange quarks were a significant
component of the nucleus, implying that a coupling between dark matter and strange quarks could lead to spin-independent scattering that could
be detected at upcoming direct detection experiments.  But more recent results from lattice QCD have suggested that the strange quark
content of the nucleon might actually be quite small.  This uncertainty is especially significant for MSSM models in the focus-point
region of parameter space; spin-independent scattering for these models proceeds through Higgs exchange, and the largest relevant Higgs-quark coupling is to the
strange quark.   These uncertainties are thus essential for determining if the focus point region is still viable or not (for example,~\cite{Buchmueller:2012hv} and~\cite{Feng:2011aa}).  We will see that
for bino-like dark matter, on the other hand, not only is it imperative to understand the quark content of nucleons, it may be that
the coupling of dark matter to first generation quarks can be the dominant contributor to scattering.

The outline of this paper is as follows.
In section II we illustrate
the relationship between the different sectors of parameter space (heavy quark, light quark, and lepton) and the observables on which we focus.
In section III we discuss the details of predicting the spin-independent scattering cross section of bino-like neutralinos with nuclei, including relevant uncertainties in the quark content of the nucleons, which are necessary to parametrize the scattering.  In section IV we describe other constraints on this
class of models arising from LHC searches, dark matter annihilation, and magnetic dipole moment measurements.
In section V we present a quantitative analysis of the sensitivity of direct detection experiments to this class of models.  We
conclude with a discussion of our results in section VI.

\section{Decoupling of three sectors}

Within the MSSM, there are a plethora of observable quantities that, if different from the Standard Model expectation, would be evidence of supersymmetry or particle dark matter.  Here, we consider the case of bino-like neutralino dark matter, which requires $M_1 \ll \mu, M_2$, where $M_1$ and $M_2$ are the bino and wino masses, respectively, and $\mu$ is the Higgs mixing parameter.
Given that no supersymmetric particles have yet been discovered at the LHC or elsewhere, we focus on the following relevant observables:
\begin{itemize}
\item {\it Higgs boson mass}: In the decoupling limit, the light $CP$-even Higgs mass in the MSSM can be approximated as
\bea
m_h^2 \approx m_Z^2 \cos^2(2\beta) + \frac{3}{4 \pi^2} \frac{m_{t}^4}{v^2}\left\{ \log \left( \frac{m_{\tilde{t}}^2}{m_t^2} \right)
+ \frac{X_t^2}{m_{\tilde{t}}^2} \left( 1-\frac{X_t^2}{12 m_{\tilde{t}}^2} \right)
\right\},
\eea
where, $m_{\tilde{t}}=\sqrt{m_{\tilde{t}_1} m_{\tilde{t}_2}}$, $\beta$ and $v$ are related to the vacuum expectation values for the Higgs doublets, $\tan \beta = v_u/v_d$ and $v_u^2 + v_d^2=v^2 = 2m_Z^2/(g^2 + g^{\prime 2})^2$, and $X_t = A_t -\mu / \tan\beta$~\cite{Carena:1995bx}.  We have neglected $\tilde{b}$ and $\tilde{\tau}$ loops, which can lead to percent-level corrections, as well as hypercharge couplings from the $D$-term potential, which result in contributions that are small compared to those due to the top Yukawa coupling.
\item {\it Dark matter annihilation cross section}:  An annihilation cross section consistent with the measured dark matter abundance within the thermal WIMP paradigm can be obtained in a variety of ways.  For example, as shown in Refs.~\cite{Pierce:2013rda} and~\cite{Fukushima:2014yia}, it is possible to achieve a relic abundance of neutralinos that would explain the dark matter in the Universe solely via $t$-channel slepton exchange in ``bulk''-type scenarios.  In this case, light sleptons with non-negligible left-right mixing are required.  Other mechanisms include coannihilations with light sleptons or squarks.  Coannihilations with light neutralinos and/or charginos are also possible, and could be accomplished for $M_1 \approx M_2$ or for a higgsino-like LSP, though each would lead to further complications in the analysis performed here.  Likewise, doping the LSP with some higgsino content would facilitate annihilations (see, eg.~\cite{Pierce:2013rda}), though, again, this would induce neutralino-nucleon scattering via Higgs exchange, a scenario we do not consider for simplicity.  Finally, the measured dark matter abundance could also be accomplished via resonant annihilations, as in the so-called Higgs funnel region~\cite{Hooper:2013qjx}.
\item {\it Dark matter-nucleon spin-independent scattering cross section}: Purely bino-like neutralinos scatter with nuclei solely via squark exchange.  In the absence of mixing between left- and right-handed squarks, this scattering is either spin-dependent or velocity-suppressed.  With mixing, however, there will also be a velocity-independent contribution to the spin-independent scattering cross section.  The cross section for spin independent elastic scattering for pure bino LSPs is therefore determined, to lowest order in velocity, only by the bino mass, $M_1$, and the relevant squark masses and mixing angles.
\end{itemize}

We are thus led to three decoupled sectors of parameter-space:
\begin{itemize}
\item {\it Heavy sector}: Choose $\mu$, heavy squark masses, and top trilinear couplings to obtain a mass for the light $CP$-even Higgs boson consistent with observations.  Also decouple the wino and gluino, as well as, possibly, some of the sleptons.
\item {\it Relic Density sector}: For a given bino mass, choose slepton masses and mixings to achieve the dark matter relic abundance.  Alternatively, the abundance may be achieved via coannihilations with squarks.
\item {\it Direct Detection sector}: For a given bino mass, neutralino-nucleon elastic scattering cross sections are determined by the light squark masses and mixings.
\end{itemize}

\section{Dark Matter Scattering with Nuclei}

In this section, we discuss the details of predicting the spin-independent elastic scattering cross section for binos on nuclei.  First, we present the framework for the calculation, assuming a model for binos interacting with quarks via squark exchange.  Next, we explore the details of the quark-nucleon matrix elements necessary to interpolate between theories of dark matter-quark scattering and limits on interactions with nuclei in detectors.  Finally, we comment on the fact that the interactions explored here are not, in general, isospin-invariant, as is often assumed for neutralino dark matter in the MSSM.

\subsection{Scattering Cross Section}
\label{sec:scatteringeqs}

\begin{figure}[hear]
\center
\includegraphics[width=0.7\textwidth]{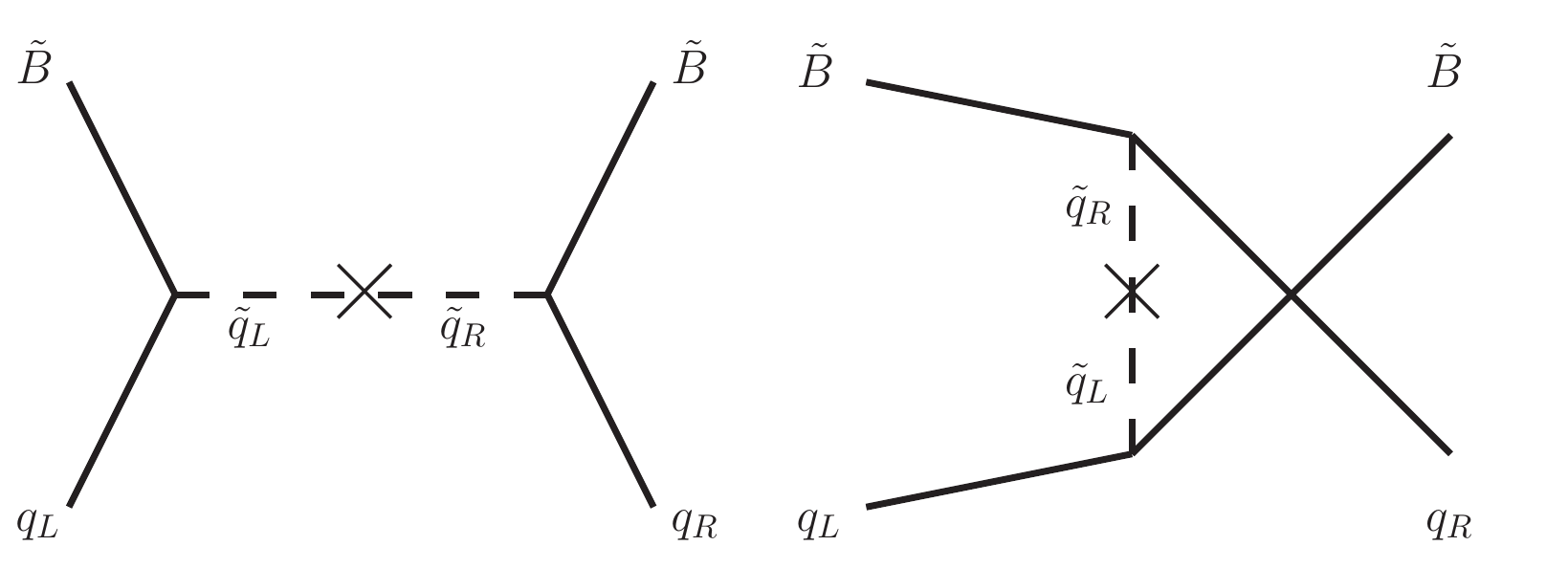}
\caption{Bino dark matter scattering with quarks via squark exchange assuming left-right mixing.}
\label{fig:sourcespec}
\end{figure}

A generic model for singlet dark matter that couples to Standard Model fermions through the exchange of charged scalars was
considered, for example, in~\cite{Fukushima:2011df}, and the results can be applied to the specific case of bino-like dark matter.
One can write the left- and right-handed squarks, $\tilde{q}_{L,R}$, as a mixture of the non-degenerate squark mass eigenstates, $\tilde{q}_{1,2}$,
\bea
\tilde{q}_L &=&  \tilde{q}_1 \cos \phi_{\tilde{q}} + \tilde{q}_2 \sin \phi_{\tilde{q}}
\nonumber\\
\tilde{q}_R &=&  - \tilde{q}_1 \sin \phi_{\tilde{q}} + \tilde{q}_2 \cos \phi_{\tilde{q}},
\eea
where $\phi_{\tilde{q}}$ is the squark mixing angle and we have dropped the potential $CP$-violating phase.
The $s$- and $u$-channel scattering of binos with quarks via squark exchange are shown in Figure~\ref{fig:sourcespec}, where the necessity of $L$-$R$ mixing is explicit.
For simplicity, and without loss of
generality, we assume $m_{\tilde{q}_{1}} \leq m_{\tilde{q}_{2}}$.
For the momentum transfer relevant for elastic scattering, the scattering interaction can be represented as a
dark matter-quark four-point contact operator.  The operator relevant for velocity-independent, spin-independent elastic scattering is
defined at the weak scale as
\bea
\cal{O} &=&
{1 \over 4} g^{\prime 2} \sin (2\phi_{\tilde{q}} ) Y_L Y_R
\left[{1 \over m_{\tilde{q}_1}^2 - m_{\tilde{\chi}}^2 } - {1 \over m_{\tilde{q}_2}^2 - m_{\tilde{\chi}}^2 } \right]
( \bar {\tilde{\chi}} \tilde{\chi} ) ( \bar q q ) \equiv \alpha_q ( \bar {\tilde{\chi}} \tilde{\chi} ) ( \bar q q ),
\label{eq:SI_contact}
\eea
where $g^\prime$ is the hypercharge coupling constant and $Y_{L,R}$ are the hypercharges of the
left- and right-handed quarks, respectively.  Note that if we had allowed for a $CP$-violating
phase, there would be additional operators that permit spin-independent scattering, but they would be velocity-suppressed.

The dark matter-nucleon spin-independent elastic scattering cross section is then given by the well-known result~\cite{SigmaSI}
\bea
\sigmaSI^N &=& {  \mu^2 \over 4 \pi  }
g^{\prime 4} Y_L^2
\left\{ \sum_{q} \sin (2\phi_{\tilde{q}}) Y_{Rq} \left[ {1 \over (m_{\tilde{q}_1}^2 - m_{\tilde{\chi}}^2) } - {1 \over (m_{\tilde{q}_2}^2 - m_{\tilde{\chi}}^2) } \right]
(B_q^N) \lambda_q \right\}^2,
\label{eq:SI_cross_section}
\eea
where $ \mu $ is the reduced mass of the DM-nucleon system, $B_q^{N}$ are the integrated
nucleon form factors for quarks $q=u,d,s$ and nucleons $N=p,n$, and the factor $\lambda_q$ accounts
for the running of the scattering operator ${\cal O}$ from the weak scale ($\sim m_Z$) to the nucleon scale $\mu$.
This factor may be expressed as $\lambda_q = m_q ( \mu) / m_q (m_Z)$, the ratio of the quark mass parameter evaluated at the
relevant scales in $\overline{MS}$ scheme~\cite{Hill:2014yxa}.
For the models we consider, we may always assume $m_p \ll m_{\tilde{\chi}}$.

In the limit of one non-decoupled light squark flavor, with $m_{\tilde{\chi}} \ll m_{\tilde q_1}$, this expression simplifies, yielding
\bea
\sigmaSI^N &\approx& (1.1 \times 10^{-8} ~\pb) Y_{Rq}^2 \sin^2 (2\phi_{\tilde{q}}) \left({1~\tev \over m_{\tilde{q}_1}} \right)^4
\left[ 1  - {m_{\tilde{q}_1}^2 \over m_{\tilde{q}_2}^2 } \right]^2 (B_q^N)^2 \lambda_q^2  \nonumber \\
	&\approx& (2.9 \times 10^{-10} ~\pb) \left(\frac{Y_{Rq}}{-1/3}\right)^2 \sin^2 (2\phi_{\tilde{q}}) \left({1~\tev \over m_{\tilde{q}_1}} \right)^4
\left[ 1  - {m_{\tilde{q}_1}^2 \over m_{\tilde{q}_2}^2 } \right]^2 \left(\frac{B_q^N}{0.5} \right)^2 \lambda_q^2 ,
\label{eq:SigmaSI_approx}
\eea
where the second line is obtained for light strange squarks, with $\sigma_{SI}^p(\tilde{s}) = 2.9 \times 10^{-10}$ pb for
maximal squark mixing, $m_{\tilde{s}_1} =1~\tev \ll m_{\tilde{s}_2}$, and a {\it minimal} reference value for $B_s^p$.  As argued in the following subsection, $B_s^p$ should not be significantly smaller than 0.5, and is potentially much larger.
In the limit $m_{\tilde{\chi}} \sim m_{\tilde q_1}$, it is more convenient to express the scattering cross-section as
\bea
\sigmaSI^N &\approx& (1.1 \times 10^{-4} ~\pb) Y_{Rq}^2 \sin^2 (2\phi_{\tilde{q}}) \left({ (100~\gev )^2 \over m_{\tilde{q}_1}^2 -m_{\tilde{\chi}}^2} \right)^2
\left[ 1  - {m_{\tilde{q}_1}^2 -m_{\tilde{\chi}}^2 \over m_{\tilde{q}_2}^2 -m_{\tilde{\chi}}^2 } \right]^2 (B_q^N)^2 \lambda_q^2 \nonumber \\
	&\approx& (2.9 \times 10^{-6} ~\pb)  \left(\frac{Y_{Rq}}{1/3}\right)^2 \sin^2 (2\phi_{\tilde{q}}) \left({ (100~\gev )^2 \over m_{\tilde{q}_1}^2 -m_{\tilde{\chi}}^2} \right)^2 \left[ 1  - {m_{\tilde{q}_1}^2 -m_{\tilde{\chi}}^2 \over m_{\tilde{q}_2}^2 -m_{\tilde{\chi}}^2 } \right]^2 \left(\frac{B_q^N}{0.5} \right)^2 \lambda_q^2,
\label{eq:SigmaSI_resonant}
\eea
where, similarly, in the second line, $\sigma_{SI}^p(\tilde{s}) = 2.9 \times 10^{-6}$ pb for $m_{\tilde{s}_1}^2-m_{\tilde{\chi}}^2 = (100~\gev)^2$,
$m_{\tilde{s}_2} \gg m_{\tilde{s}_1}$.  These approximate expressions indicate that the spin-independent scattering cross sections due to squark exchange are in the range to be explored by direct dark matter searches.  Furthermore, the cross sections for scattering via the exchange of any light squark may be obtained by rescaling Equations~\ref{eq:SigmaSI_approx} or~\ref{eq:SigmaSI_resonant} with the appropriate hypercharge, nucleon form factor, and squark mixing angle.  We note that the following analysis makes use of the more exact expression, Equation~\ref{eq:SI_cross_section}.

\subsection{Quark Content of the Nucleon}

The integrated nucleon form factors for light quarks are just the quark nucleon matrix elements,
\bea
B_q^{N} =   \langle N | \bar{q} q | N \rangle = \frac{m_N}{m_q}f_q^{N},
\eea
where we note that the matrix elements are also proportional to the form factors $f_q^{N}$, with the constant of proportionality being the ratio of the nucleon mass, $m_N$, to the quark mass parameter, $m_q$ (which will be evaluated at $2~\gev$ in $\overline{MS}$ scheme).
Proton and neutron matrix elements are related by
\bea
B_u^{n} = B_d^{p}, B_d^{n} = B_u^{p}, \textrm{ and } B_s^{n} = B_s^{p}.
\eea

Numerical values for each $B_q^{N}$ may be determined from a variety of techniques.  It is generally more convenient to consider the parameters $\Sigma_{\pi N}$, $\sigma_0$, and $z$:
\bea
&\Sigma_{\pi N} \equiv \frac{\displaystyle m_u + m_d}{\displaystyle 2} \left( B_u^{N} + B_d^{N} \right), \\ \nonumber
&\sigma_0 \equiv \frac{\displaystyle m_u+m_d}{\displaystyle 2} \left( B_u^{N}+B_d^{N}-2B_s^{N} \right), \\
&z \equiv {\displaystyle B_u^p-B_s^p \over \displaystyle B_d^p-B_s^p}. \nonumber
\eea
We see that the nucleon form factors, which often appear in calculations of the spin-independent scattering of WIMPs with nuclei,
 can then be expressed as
\bea
f_{u}^{N} &=&  { 2 \Sigma_{\pi N} \over m_N \left(1 + {m_d \over m_u}\right) \left(1 + {B_d^{N} \over B_u^{N}}\right)}
\nonumber\\
f_{d}^{N} &=&  { 2 \Sigma_{\pi N} \over m_N \left(1 + {m_u \over m_d}\right) \left(1 + {B_u^{N} \over B_d^{N}}\right)}
\nonumber\\
f_{s}^{N} &=&  { \left( {m_s \over m_d} \right) \Sigma_{\pi N} \, y \over m_N \left(1 + {m_d \over m_u}\right)},
\eea
where $y$ is typically referred to as the strangeness content of the nucleon,
\bea
y\equiv 1-\frac{\sigma_0}{\Sigma_{\pi N}} = 2\frac{B_s^{N}}{B_u^{N} + B_d^{N}}.
\eea
The parameter $z$ is calculated purely from the baryon octet mass differences, with the result $z=1.49$ (this result has small errors relative to the uncertainty in $\sigma_0$ and $\Sigma_{\pi N}$ as discussed below)~\cite{Cheng:1988im}.
At the present moment, the largest drivers of uncertainty in the calculation of neutralino-nucleon elastic scattering cross sections
are the strangeness content of the nucleon~\cite{Draper:2013cka}, the measurement of $\Sigma_{\pi N}$, and the measurement of the
quark masses.

The value of $\Sigma_{\pi N}$ can be determined from pion-nucleon scattering data, but with quite large uncertainties.
Recent analyses find $\Sigma_{\pi N} = 64 \pm 8$ MeV~\cite{Pavan:2001wz} and $\Sigma_{\pi N} = 59 \pm 7$ MeV~\cite{Alarcon:2011zs};
these values are larger by more than $2\sigma$ than those found in previous analyses~\cite{Gasser:1990ce}.

$\sigma_0$ can either be fit from the baryon masses in chiral perturbation theory, or inferred from lattice QCD studies.
The former method, when applied to the baryon octet masses at ${\cal O}(p^2)$ at tree-level in chiral perturbation theory,
yields $\sigma_0 = 27$ MeV and a very large strangeness content such that $f_{s}^{N}\gg f_{u,d}^{N}$.
On the other hand, recent lattice studies favor $f_{s}^{N}\approx f_{u,d}^{N}$, but predict a small value for $\Sigma_{\pi N}$ that is in tension with recent determinations~\cite{Thomas:2012tg}.  Moreover,
recent studies show that the chiral perturbation theory result may also be consistent
with much larger values of $\sigma_0$~\cite{Borasoy:1996bx, Alarcon:2012nr} and with $f_{s}^{N}\approx f_{u,d}^{N}$, if
one includes higher orders in the momentum expansion and includes the baryon decouplet.  In this case, it is possible to reconcile small strangeness content with the large value of $\Sigma_{\pi N} \approx 60$ MeV.

As discussed in~\cite{Ellis:2008hf}, it is not only the values of the quantities $\Sigma_{\pi N}$ and $\sigma_0$ that affect the predicted scattering cross sections, but also their difference: for $\sigma_0 \approx \Sigma_{\pi N}$ ($y \approx 0$), the strangeness content of the nucleon becomes negligible, while for larger differences, the strangeness content dominates.  This is immediately obvious if we consider the integrated nucleon form factors,
\bea
B_u^p = B_d^n &=& \tilde{\Sigma}_{\pi N} + \tilde{\sigma}_{0}\left( \frac{z-1}{z+1} \right)
	= \tilde{\Sigma}_{\pi N}\left[ 1 +(1-y)\left( \frac{z-1}{z+1} \right) \right]  \nonumber \\
B_d^p = B_u^n &=& \tilde{\Sigma}_{\pi N} - \tilde{\sigma}_{0}\left( \frac{z-1}{z+1} \right)
	=  \tilde{\Sigma}_{\pi N}\left[ 1 -(1-y)\left( \frac{z-1}{z+1} \right) \right] \\
B_s^p = B_s^n &=& \tilde{\Sigma}_{\pi N}-\tilde{\sigma}_{0} =  \tilde{\Sigma}_{\pi N}\,y, \nonumber
\eea
where $\tilde{\Sigma}_{\pi N}= \Sigma_{\pi N}/(m_u+m_d)$, $\tilde{\sigma}_{0} = \sigma_0 /(m_u+m_d)$. We see that even if strange squarks are heavy and contribute negligibly to dark matter-nucleon scattering, the
uncertainty in the dark matter-nucleon scattering cross section is still influenced by the strange quark content of the nucleon.
The values of $B_q^{N}$ are given in Table~\ref{tab:Buds} for the limiting reference cases $y\rightarrow 0$ and $y\rightarrow1$, as well as the benchmark case $y=0.06$ which yields $B_s^N=0.5$ for the central value $\Sigma_{\pi N}=59$ MeV.  We note that this corresponds to $f_s^N =0.0567$, which is, for example, 27\% larger than the value $f_s^N =0.0447$ set as the default in micrOMEGAS versions 3.0 to 3.5.5~\cite{Belanger:2013oya}.

The final source of uncertainty that we discuss is that in the masses of the light quarks, which cannot be neglected (in contrast to the example
studied in~\cite{Ellis:2008hf}).
A pure bino LSP will only scatter via squark exchange, and scattering will be spin-independent in the non-relativistic limit
only if there is $L$-$R$ mixing in the squark sector; this is the dominant process we consider.
However, typically, the $L$-$R$ mixing is taken to be negligible for light squarks in the MSSM ($\phi_{\tilde{q}} \approx 0$), leading to a nearly-diagonal mixing matrix (details are included in the Appendix; specifically, the mixing matrix given in eq.~\ref{eq:SfermionMassMatrix}).
In this case, the higgsino content of the LSP dominates spin-independent scattering, via Higgs exchange.  As discussed in~\cite{Ellis:2008hf}, one then finds
$\sigma_{SI}^{N} \propto (m_q B_q^{N})^2$.  Since $B_q^{N} \propto 1/(m_u+m_d)$, the entire scattering cross section depends only on ratios of the light quark masses, which are fairly well constrained~\cite{Leutwyler:1996qg}.  Here, we focus specifically on spin-independent scattering via squark exchange,
for which $\sigma_{SI}^{N} \propto (B_q^{N})^2 \propto 1/(m_u+m_d)^2$; $\sigma_{SI}^{N}$ clearly depends on the absolute masses of the up and down quark.  In Table~\ref{tab:Buds} and throughout this analysis, we consider only the central values for the light quark masses~\cite{Agashe:2014kda} defined at
$2~\gev$ using $\overline{MS}$ scheme,
\bea
m_u = 2.3^{+0.7}_{-0.5} \textrm{ MeV} \,\,\, \textrm{ and } \,\,\,
m_d = 4.8^{+0.5}_{-0.3} \textrm{ MeV},
\eea
and note that all $B_q^{N}$ scale as $1/(m_u+m_d)$, which is an additional source of uncertainty in the ability of experiments to constrain these models\footnote{We note also that lattice QCD calculations~\cite{Davies:2009ih} yield more precise values for the quark masses than the results of measurements presented in~\cite{Agashe:2014kda}.}.  In the figures presented in Section~\ref{sec:analysis}, we display only a cross section band related to the uncertainty in $\Sigma_{\pi N}$ and the strangeness content of the nucleon.

\begin{center}
\begin{table}
%\begin{tabular*}{.6\textwidth}{@{\extracolsep{\fill} } | c || c c | c | c c |}
\begin{tabular}{ | c || c | c || c || c | c |}
\hline
 	& \multicolumn{2}{c||}{$y\rightarrow0$} & $y=0.06$ & \multicolumn{2}{c|}{$y\rightarrow1$} \\
\hline
$B_u^p = B_d^n$ & $\frac{2z}{1+z}\, \tilde{\Sigma}_{\pi N} $ & 9.95 (7.59, 12.2)  & 9.85 (7.51, 12.1) & $\tilde{\Sigma}_{\pi N}$ & 8.31 (6.34, 10.3) \\
$B_d^p = B_u^n$ & $ \frac{2}{1+z}\, \tilde{\Sigma}_{\pi N}$ & 6.67 (5.09, 8.38) & 6.77 (5.17, 8.46) & $\tilde{\Sigma}_{\pi N}$ & 8.31 (6.34, 10.3)  \\
$B_s^p = B_s^n$ & $0$ & 0 & 0.499 (0.380, 0.617)& $\tilde{\Sigma}_{\pi N}$ & 8.31 (6.34, 10.3)  \\
\hline
\end{tabular}
\caption{Values for the integrated nucleon form factors, $B_q^{N}$, for some benchmark cases. Numerical values are unitless	and assume the central value for $\Sigma_{\pi N}$ of 59 MeV, with the numbers in parentheses indicating the $2\sigma$ range for $\Sigma_{\pi N}$ (45 MeV, 73 MeV)~\cite{Alarcon:2011zs}.  The limit $y = 0.06$ is the benchmark case that yields $B_s^N=0.5$ for the central value $\Sigma_{\pi N} =59$ MeV.  Here and throughout we take $m_u = 2.3$ MeV and $m_d = 4.8$ MeV~\cite{Agashe:2014kda}.  While the uncertainty in the light quark masses is non-negligible, each $B_q^{N}$ scales as $1/(m_u+m_d)$ as discussed in the text.  We therefore focus on the uncertainty in $\Sigma_{\pi N} $ and the strangeness content of the nucleon. }
\label{tab:Buds}
\end{table}
\end{center}

Finally, we note that if the superpartners of the heavy quarks ($c$, $b$ and $t$) exhibit left-right mixing,
they can also potentially contribute to bino-nucleon spin-independent scattering via a heavy quark loop, which couples to the
gluon content of the nucleon.  The effective integrated nucleon form factors for the heavy quarks can be determined from anomaly
considerations~\cite{Shifman:1978zn}, and are given by
\bea
B_{Q=c,b,t}^N &=& {2 \over 27} {m_N \over m_Q} f_g^N
\nonumber\\
f_g^N &=& 1 - \sum_{q=u,d,s} f_q^N
\eea
Since $B_Q^N \propto m_Q^{-1}$, these contributions are expected to be subleading unless the light quark
superpartners are either heavier or have suppressed left-right mixing.  We also point out that since one might expect the
strange quark content of the nucleon to be at least as large as that of the heavy quarks (i.e.~ $f_s^N > f_c^N$), one would expect
$B_s^N \gtrsim 0.5$, which is the reference value used in Equations~\ref{eq:SigmaSI_approx} and~\ref{eq:SigmaSI_resonant} .

\subsection{Isospin Violation}

In this scenario, dark matter interactions are generically isospin-violating~\cite{Kurylov:2003ra,Giuliani:2005my,Chang:2010yk,Kang:2010mh,Feng:2011vu,Feng:2013vod}.
Isospin violation in spin-independent scattering
is typically parameterized in terms of $f_{p,n}$, the coupling of dark matter to protons and neutrons.  In particular,
$\sigmaSI^n / \sigmaSI^p = (f_n / f_p)^2$.  In this scenario, this ratio of couplings can be expressed as
\bea
{f_n \over f_p} &=& {  \sum_{q} \sin (2\phi_{\tilde{q}}) Y_{Rq} \left[ {1 \over (m_{\tilde{q}_1}^2 - m_{\tilde{\chi}}^2) } - {1 \over (m_{\tilde{q}_2}^2 - m_{\tilde{\chi}}^2) } \right]
(B_q^n) \lambda_q
\over
\sum_{q} \sin (2\phi_{\tilde{q}}) Y_{Rq} \left[ {1 \over (m_{\tilde{q}_1}^2 - m_{\tilde{\chi}}^2) } - {1 \over (m_{\tilde{q}_2}^2 - m_{\tilde{\chi}}^2) } \right] (B_q^p) \lambda_q }.
\eea
Because $B_s^p \sim B_s^n$, bino spin-independent scattering will be largely isospin-invariant if
it is dominated by strange squark exchange.
Generically, however, $f_n / f_p$ may assume any value, including negative values.
In fact, even if up and down squarks are mass degenerate and have the same mixing angle,
bino scattering will be isospin-violating because $Y_{Ru} \neq Y_{Rd}$.
It is thus worth noting that even a
bino-like LSP of the MSSM can be an example of isospin-violating dark matter (IVDM).

In the limit where spin-independent scattering is dominated by the exchange of one squark flavor, the
above expression simplifies to $f_n / f_p \sim B_q^n / B_q^p$.  If scattering is dominated by the exchange of up,
down, or strange squarks, and the strangeness content is relatively small, $f_n / f_p \approx 2/3, 3/2, 1$, respectively (see, for example,~\cite{Hill:2014yxa}).

\section{Other Signals and Constraints}

If one introduces large left-right sfermion mixing, one may potentially generate new vacua
under which $U(1)_{EM}$ or $SU(3)_{QCD}$ are broken.  Severe constraints on the mixing angle
arise if one requires that the neutral vacuum is a global minimum.  But if one only requires
that the neutral vacuum is metastable with a lifetime of order the age of the universe, then the
constraints are considerably weaker (see, for example,~\cite{Hisano:2010re}).  For the masses
considered here, even maximal mixing is consistent with the metastability constraint.

There are several other potential signals and constraints for this scenario, which we address.

\subsection{Bounds on Squark Masses from the LHC}

Exclusion limits for direct squark production with decoupled gluinos at LHC require $m_{\tilde{q}} \gsim { \cal O } ( \tev)$ assuming eight
degenerate light flavor squarks and $m_{\tilde{\chi}} \sim {\cal O } ( 100 \gev )$~\cite{Aad:2014wea}. If we allow for one non-degenerate light flavor
squark then the mass constraints are significantly weakened to $m_{\tilde{q}_1} \gsim { \cal O } ( 500 \gev)$. In the limit of decoupled gluinos,
light flavor squarks are produced primarily through an $s$-channel gluon, resulting in similar production cross sections for first and second
generation squarks.  As the gluino mass decreases, $t$-channel gluino exchange will start to dominate squark production, and the first generation
squark production cross section will be enhanced relative to second generation squarks due to the parton distribution functions (PDFs). Thus,
for $m_{\tilde{g}} \sim \cal{O} ( \tev) $, the exclusion limits for a non-degenerate first generation squark are roughly twice as strong as the exclusion limits
for a non-degenerate second generation squark~\cite{Mahbubani:2012qq}.

\subsection{Dark Matter Annihilation}

If light squarks exhibit significant left-right mixing, then bino dark matter
can annihilate via $t$-channel squark exchange ($\tilde{\chi}\tilde{\chi} \rightarrow \bar q q$), with neither
$p$-wave nor chirality suppression.  This scenario was considered in~\cite{Fukushima:2014yia}, for
example, in the context of light sleptons; the relevant annihilation cross sections for light
squarks should be rescaled by a factor $<1$, which accounts for the squark hypercharges and the final state
color factor.
In~\cite{Fukushima:2014yia},
it was shown that in order to obtain $\langle \sigma_{\tilde{\chi}\tilde{\chi}  \rightarrow \bar f f} v \rangle \sim 1~\pb$, one
would need sfermions as light as $\sim 150~\gev$.  Given current constraints on squark masses, the
process $\tilde{\chi}\tilde{\chi} \rightarrow \bar q q$ can only play a subleading role in dark matter freeze out.
However, as shown in~\cite{Fukushima:2014yia}, slepton
masses are much less constrained, allowing the annihilation process $\tilde{\chi}\tilde{\chi}  \rightarrow \bar \ell \ell$ to
sufficiently deplete the bino relic density.

However, in the region of parameter space where $m_{\tilde{\chi}} \sim m_{\tilde q_1}$, the bino-squark co-annihilation
process $\tilde{\chi} \tilde q_1 \rightarrow q (\gamma, Z, g, h) $ can be significant in the early universe.
Moreover, the spin-independent scattering cross section receives a resonant enhancement in this region of parameter space
(see eq.~\ref{eq:SigmaSI_resonant}).  This ``light squark co-annihilation" region is thus correlated with
interesting signals at direct detection experiments.

\subsection{Constraints on Dipole Moments of Quarks}

If the light squarks exhibit left-right mixing, then there can be a significant contribution to the
quark magnetic dipole moment arising from a bino-squark one-loop diagram (a similar effect has been
discussed in the context of light sleptons~\cite{Cheung:2009fc,Fukushima:2013efa,Fukushima:2014yia}).
The contribution to the magnetic moment in the
$m_q \rightarrow 0$ limit is
\bea
{\Delta a \over m_q } &\sim& { m_{\tilde{\chi}} \over 8 \pi^2 m_{\tilde{q}_1}^2} g^2 Y_L Y_{Rq} \sin (2\phi_{\tilde{q}})
\left[ {1 \over 2 (1-r_1)^2 } \left( 1 + r_1 + { 2 r_1 \ln r_1 \over 1 - r_1} \right)\right] - (m_{\tilde{q}_1} \leftrightarrow m_{\tilde{q}_2}),
\label{eq:DipoleMoment}
\eea
where $ r_i =m_{\tilde{\chi}}^2 / m_{\tilde{q}_i}^2$.

In the limit $r_i \sim 0$, $m_{\tilde{q}_1} \ll m_{\tilde{q}_2}$ limit,
we find
\bea
\alpha_q &\sim&   {16\pi^2 \over m_{\tilde{\chi}}} {\Delta a_q \over m_{q} }
\nonumber\\
\sigmaSI^N &\sim& \left({16\pi^2 \over m_{\tilde{\chi}}} \right)^2 { 4 \mu^2 \over \pi }
\left( \sum_{q} { \Delta a_q \over m_{q} } B_q^N \right)^2
\nonumber\\
&\sim& (1.1 \times 10^{9}~\pb \gev^2) \left( \sum_{q}  {\Delta a_q \over m_{q} } {B_q^N \over 0.5} \right)^2
\left(m_{\tilde{\chi}} \over 50~\gev \right)^{-2}.
\label{eq:SigmaSI_DipoleMoment}
\eea
Thus current constraints from direct detection experiments already rule out models for which
$\Delta a_q (\gev / m_{q}) \gtrsim 10^{-9}$ for $m_{\tilde{\chi}} = 50$ GeV.

The magnetic dipole moments of light current quarks are constrained by LEP data~\cite{Escribano:1993xr}, because
new physics that can contribute to an anomalous helicity-mixing term in the
$\bar f f \gamma$ vertex can also contribute to an anomalous helicity-mixing
term in the $\bar f f Z$ vertex, leading to a new contribution to the width of the
$Z$.  These bounds depend on the precise coupling, but the tightest bounds are roughly
$\Delta a_{u,d} \lesssim 10^{-5}$, $\Delta a_s \lesssim 10^{-3}$.  Applying Eq.~\ref{eq:SigmaSI_DipoleMoment}, where
we take $m_q$ to be a current quark mass, we see that these bounds are also not very constraining
for $\sigmaSI^N$ relevant for direct detection.

The magnetic dipole moments of light constituent quarks can also be constrained by
measurements of the magnetic dipole moments of baryons.  In the static quark model,
the magnetic moments of the proton and neutron, $g_p$ and $g_n$, can be expressed
in terms of the quantities $\Delta a_q (m_N / m_{q})$.  However, a minimum level of uncertainty
in this expression can be roughly estimated by the deviation of the measured
baryon dipole moments from the predicted ratio $g_n / g_p = -2/3$.  This uncertainty is thus at
least greater than ${\cal O}(10^{-2})$, implying that the baryon moments do not significantly constrain
the spin-independent scattering cross sections relevant for this analysis.

A bino-squark loop diagram also contributes to the chromomagnetic
dipole moment of the quark.  This contribution is also given by eq.~\ref{eq:DipoleMoment},
up to a QCD group theory factor.  The anomalous chromomagnetic moment of the current quarks
can be bounded by the LHC at $\Delta a \lesssim 10^{-5}$~\cite{Hesari:2014hva}, assuming a contact
interaction.  This bound should be viewed somewhat heuristically, as the contact approximation
may not be valid at LHC energies, if the squarks are not sufficiently heavy.  But in any case, these bounds
are thus also superseded by current bounds from direct detection experiments.

\section{Analysis}
\label{sec:analysis}

We now consider the sensitivity of direct detection experiments to the class of models
considered here.
First, we focus on strange squarks as the mediators of bino-nucleon scattering.  At the end of this section, we will return to the possibility of up and/or down squarks as the mediators.
As the effective operator is defined at the weak scale, while the integrated nucleon form factors are evaluated using
quark masses defined at $2~\gev$, we find that $\lambda_q = m_q (2~\gev) / m_q (m_Z)$, where the quark mass parameters
are evaluated in $\overline{MS}$ scheme.  Values for these running quark masses may be found in~\cite{Xing:2007fb}.

We will focus on the parameter space in which the operator given in eq.~\ref{eq:SI_contact} provides the leading
contribution to spin-independent scattering.  There are competing higher-dimension dark matter/quark operators, but the resulting
contributions to the scattering matrix element are velocity-suppressed and will be subleading provided $\phi_{\tilde q}
\gtrsim 10^{-3}$.  Moreover, there are dark matter/gluon operators which arise from loops of heavy
quarks ($Q=c, b, t$).  These contributions are also subleading provided we consider the region of parameter
space $\alpha_q B_q^N \gg \alpha_Q B_Q^N$; this constraint will be easily satisfied provided the heavy squarks are decoupled,
or have very weak left-right mixing.

\begin{figure}[hear]
\center
\includegraphics[width=0.7\textwidth]{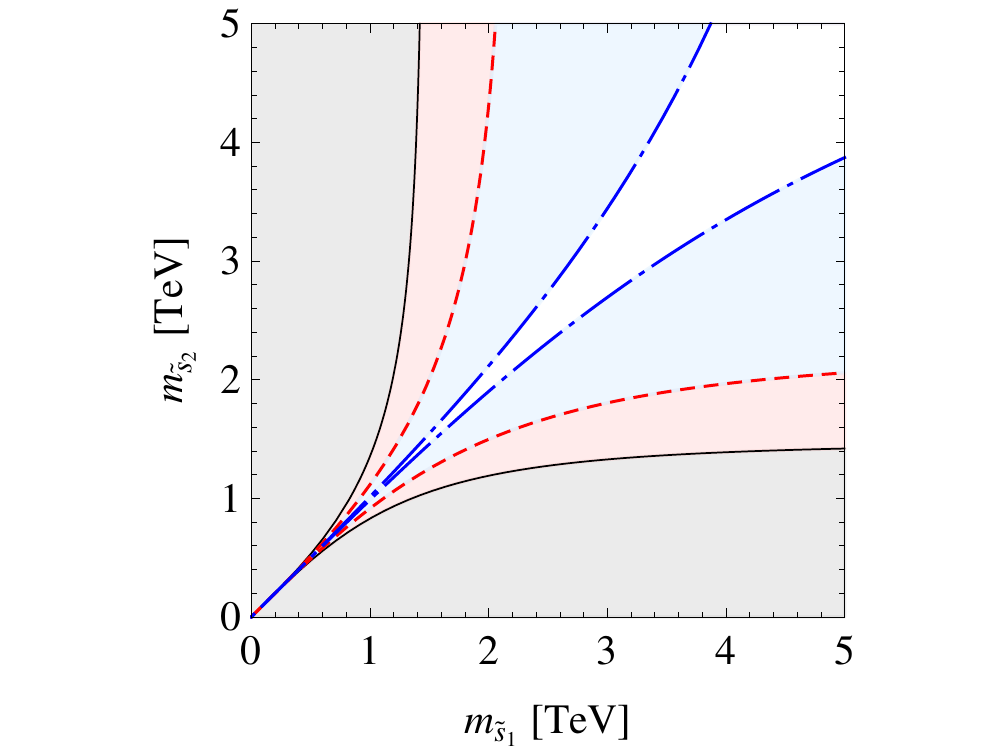}
\caption{Direct detection sensitivity plots in the $(m_{\tilde s_1}, m_{\tilde s_2})$ plane for bino dark matter
with $m_{\tilde{\chi}} = 50~\gev$.  We assume that only the strange squarks are light, with maximal left-right squark mixing,
and take the minimal reference value $B_s^N =0.5$.  The grey region (between the axes and the solid lines) is ruled out by current LUX data~\cite{LUX}, while the red region (bewtween the solid and dashed lines) could be probed by LUX with 300 days of data, and the
blue region (between the dashed and dot-dashed lines) could be probed by LZ-7~\cite{LUXproj}.  }
\label{fig:m1-m2-plane}
\end{figure}

If strange squarks are relatively light, while all other squarks are decoupled, bino-nucleon scattering is isospin invariant, so one can directly compare the predicted scattering cross sections to the published sensitivities of direct dark matter searches.
In Figure~\ref{fig:m1-m2-plane}, we plot sensitivity contours for LUX,
in the $(m_{\tilde s_1}, m_{\tilde s_2})$ plane, for $m_{\tilde{\chi}} = 50~\gev$ and maximal left-right
squark mixing, assuming $B_s^N = 0.5$.  The grey region is ruled out by
current LUX data~\cite{LUX}, while the red region could be probed by LUX with 300 days of data, and the
blue region could be probed by LZ-7~\cite{LUXproj}\footnote{All limits and sensitivities are reported at the 90\% confidence level, as is the standard practice for direct dark matter searches.}.
Note that if $m_{\tilde{\chi}} \ll m_{\tilde q_{1,2}}$,  $\sigmaSI^{N} \propto \sin^2 (2\phi_{\tilde{q}}) Y_{Rq}^2 (B_q^{N})^2  \lambda_q^2 
m_{\tilde q}^{-4}$;
one can determine the LUX sensitivity for a different mixing angle or integrated nucleon form factor by rescaling $m_{\tilde q_{1,2}}$
appropriately.  Similarly, the LUX sensitivity for different squark flavors is easily obtained by rescaling $m_{\tilde q_{1,2}}$ for the relevant choice of parameter values.

\begin{figure}[hear]
\center
\includegraphics[width=0.7\textwidth]{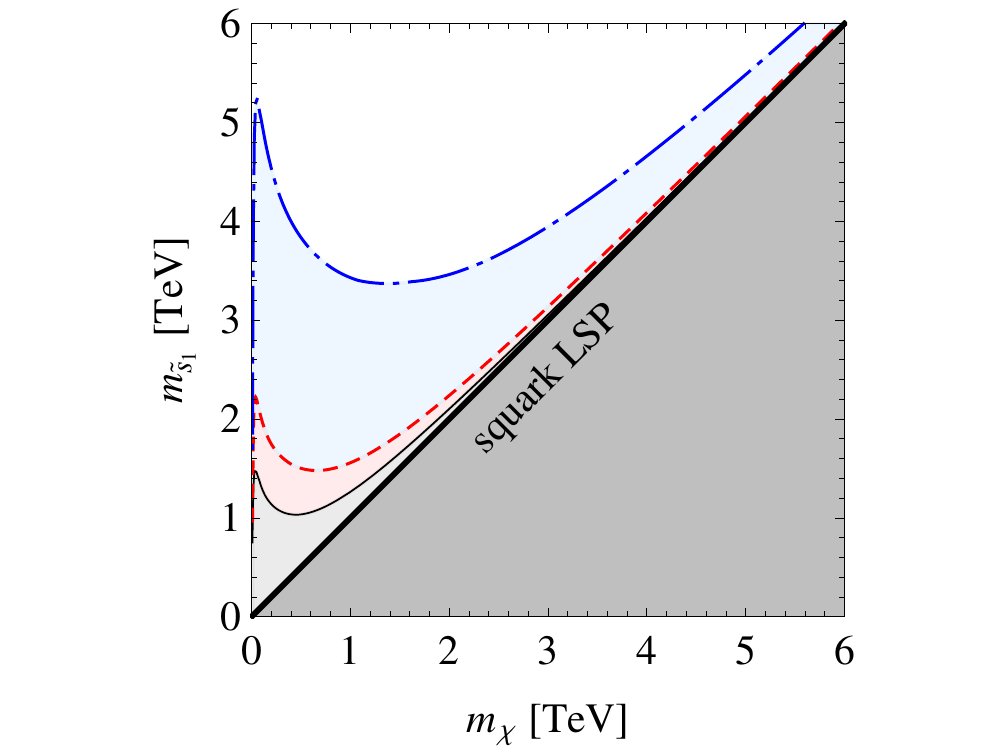}
\caption{Direct detection sensitivity plots in the $(m_{\tilde{\chi}}, m_{\tilde s_1})$ plane for bino dark matter
with $m_{\tilde s_2} = 10~\tev$.  We assume that only the strange squarks are light, with maximal left-right squark mixing,
and take the minimal reference value $B_s^N =0.5$.  The grey region (between the solid line and squark LSP line) is ruled out by
current LUX data~\cite{LUX}, while the red region (between the solid and dashed lines) could be probed by LUX with 300 days of data, and the
blue region (between the dashed and dot-dashed lines) could be probed by LZ-7~\cite{LUXproj}.  Note, only the region $m_{\tilde{\chi}} < m_{\tilde s_1}$ is physically relevant.}
\label{fig:mchi-m1-plane}
\end{figure}

In Figure~\ref{fig:mchi-m1-plane}, we plot sensitivity contours for direct detection experiments in the
$(m_{\tilde{\chi}}, m_{\tilde s_1})$ plane for $m_{\tilde s_2} =10$ TeV, $\sin (2\phi_s)=1$,
and $B_s^N =0.5$.
The color scheme is the same as in Figure~\ref{fig:m1-m2-plane}.
Only the region $m_{\tilde s_1} > m_{\tilde{\chi}}$ is relevant; otherwise the lightest supersymmetric particle would be a squark.
Even for the somewhat conservative value of $y\approx0.06$ that leads to $B_s^N =0.5$, direct dark matter searches will be sensitive to models in which the lightest colored superpartners are multi-TeV squarks with masses well beyond the reach of the LHC operating at a center of mass energy of 14 TeV (LHC-14)~\cite{CMSfuture}.
It is interesting to note that direct detection experiments are sensitive even to models
with $m_{\tilde{\chi}}, m_{\tilde s_1} \sim {\cal O}({\rm several }\tev)$, provided the bino and lightest squark are sufficiently degenerate. This reflects the resonant enhancement in the scattering cross section.   For $m_{\tilde{\chi}} \approx m_{\tilde{q}_1}$, it is important to note that
although the scattering cross section formally diverges at the point of exact degeneracy in the massless quark limit,
the physical event rate exhibits no such divergence.  In the expression for the scattering cross section (Eq.~\ref{eq:SI_cross_section}),
we have assumed that the momentum transfer is negligible; in the exact expression, one instead finds
$d\sigma_{\rm SI}^N / dE_R \propto E_R^{-2}$ in the degenerate limit, where $E_R$ is the recoil energy.  Any physical direct detection experiment is insensitive
to scatters with $E_R < E_{th}$, where $E_{th}$ is the experimental threshold; the actual event rate is thus finite.

Finally, Figure~\ref{fig:mchi-m1-plane} demonstrates that for $m_{\tilde{s}_2}=10$ TeV,
even for $m_{\tilde{\chi}} \ll m_{\tilde{s}_1}$, LZ-7 will be sensitive to all models with maximal mixing for which $m_{\tilde{s}_1} \lesssim 3.2$ TeV.  If the strangeness content of the nucleon is larger than $y\approx0.06$ (that is, if $\sigma_0$ is significantly smaller than $\Sigma_{\pi N}$), then LZ-7 will be sensitive to models in which the lightest colored superpartner is well beyond the reach of LHC-14, irrespective of the mass of the dark matter particle.  For example, for $B_s^N=0.76$ (corresponding to a strangeness content of $y\approx 0.1$ for $\Sigma_{\pi N} = 59$ MeV), the LZ-7 sensitivity region in Fig.~\ref{fig:mchi-m1-plane} would extent beyond 3 TeV for all dark matter masses $m_{\tilde{\chi}} \gtrsim 15$ GeV.  Moreover, the reach in $m_{\tilde{s}_1}$ improves for heavier $m_{\tilde{s}_2}$.
If, on the other hand,  the squarks are degenerate, it is clear from Eq.~\ref{eq:SI_contact} or~\ref{eq:SI_cross_section} that the scattering cross section due to squark exchange will vanish.

\begin{figure}[hear]
\center
\includegraphics[width=0.7\textwidth]{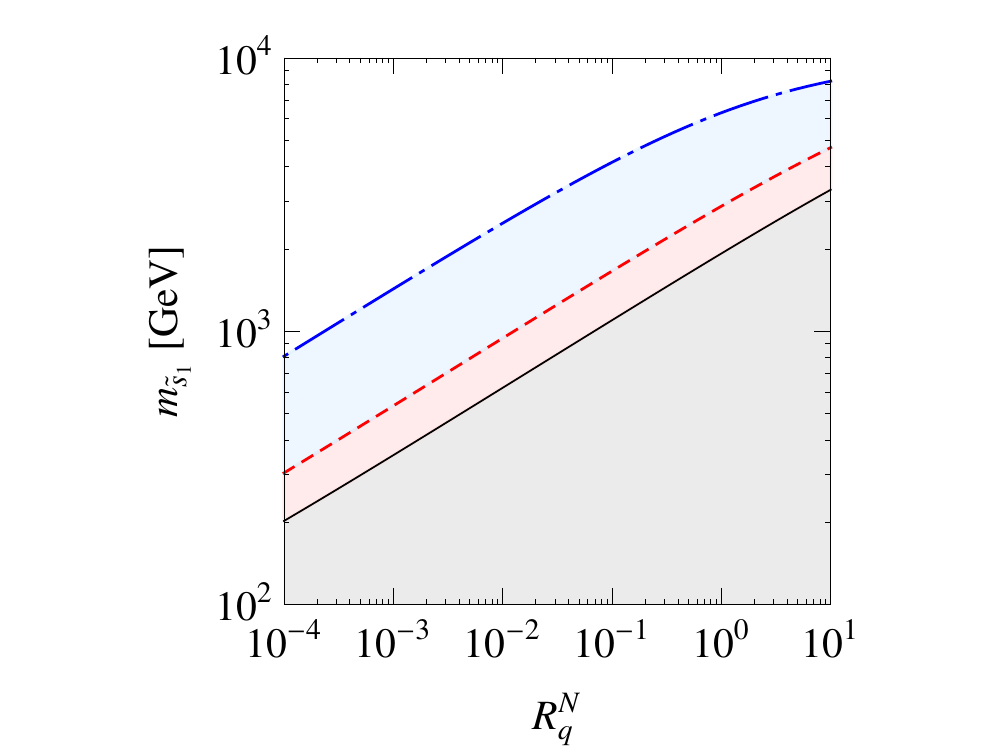}
\caption{Direct detection sensitivity plots in the $(R_s^N, m_{\tilde s_1})$ plane for bino dark matter
with $R_q^N \equiv Y_{Rq}^2 \sin^2 (2\phi_{\tilde{q}}) (B_q^N)^2 \lambda_q^2$, $m_{\tilde s_2} = 10~\tev$, $m_{\tilde{\chi}} = 50~\gev$.
We assume that only the strange squarks are light.  The grey region (between the plot frame and the solid line) is ruled out by
current LUX data~\cite{LUX}, while the red region (between the solid and dashed lines) could be probed by LUX with 300 days of data, and the
blue region (between the dashed and dot-dashed lines) could be probed by LZ-7~\cite{LUXproj}.  }
\label{fig:scale-m1-plane}
\end{figure}

In Figure \ref{fig:scale-m1-plane}, we plot sensitivity curves for direct detection experiments in the
$(R_s^{N} , m_{\tilde s_1})$ plane, where
$R_q^{N} \equiv Y_{Rq}^2 \sin^2 (2\phi_{\tilde{q}}) (B_q^{N})^2 \lambda_q^2$ and we assume
$m_{\tilde{\chi}} = 50~\gev$, $m_{\tilde s_2} =10~\tev$.
The color scheme is the same as in Figure~\ref{fig:m1-m2-plane}.  We see that next generation direct detection experiments will
be sensitive to models with squark masses that are allowed by LHC
constraints and with mixing angles as small as $\phi_{\tilde{s}} \sim 0.01$ in the $m_{\tilde{\chi}} \ll m_{\tilde s_1}$ regime, even
assuming $B_s^N = 0.5$.  If
$m_{\tilde{\chi}} \sim m_{\tilde s_1}$, then direct detection experiments may be sensitive to even smaller
mixing angles.  Note however that for $\phi_{\tilde{q}} \lesssim 10^{-3}$, one would expect velocity-suppressed
contributions to the spin-independent scattering cross section (which we have ignored) to become comparable to those
arising from left-right squark mixing.

\begin{figure}[hear]
\center
\includegraphics[width=0.7\textwidth]{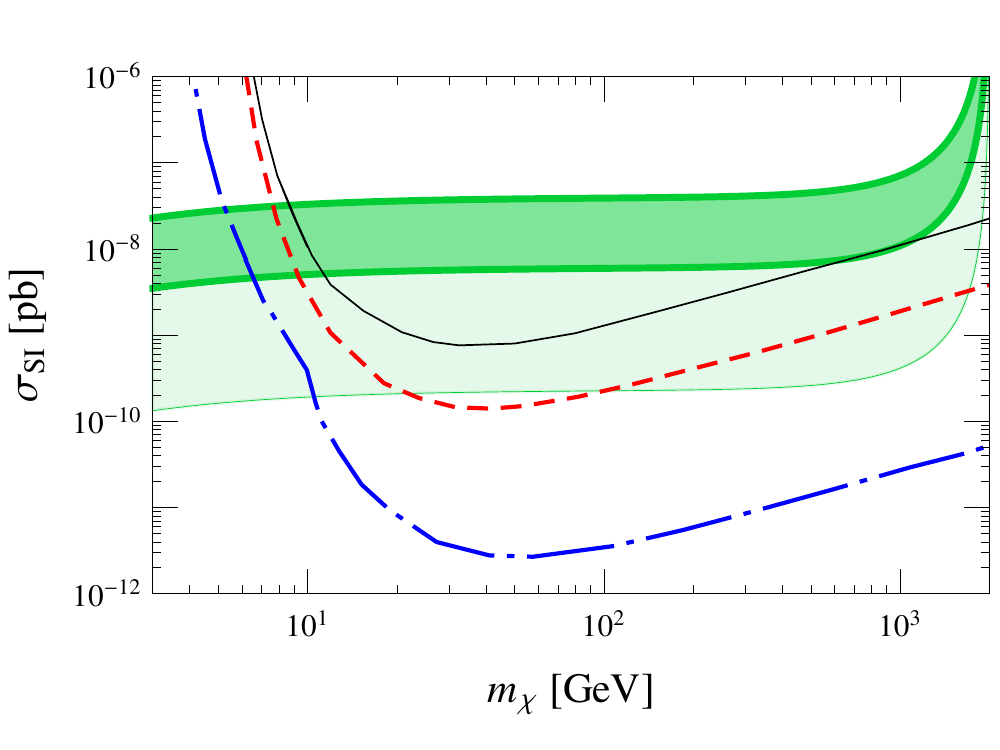}
\caption{Direct detection sensitivity plots in the $(m_{\tilde{\chi}}, \sigmaSI^N)$ plane for bino dark matter
with $m_{\tilde s_1} =2~\tev$, and with all other squarks assumed to be heavy.  We assume maximal left-right squark mixing.
The black line is the bound from
current LUX data~\cite{LUX}, while the dashed red line is the sensitivity curve for LUX with 300 days of data, and the
blue dot-dashed curve could be probed by LZ-7~\cite{LUXproj}.
The green bands indicate the uncertainty in the scattering cross section as a function of the strangeness content
of the nucleon.  The dark green band indicates the predicted SI-scattering cross section
for $\sigma_0 =27~\mev$ and allowing the full $2\sigma$ range for $\Sigma_{\pi N}$ of 45 MeV to 73 MeV~\cite{Alarcon:2011zs}.
The light green band indicates the predicted SI-scattering cross section if one decreases the strangeness content
further, to a minimum of $B_s^N =0.5$.  }
\label{fig:mchi-sigma-plane}
\end{figure}

In Figure \ref{fig:mchi-sigma-plane}, we plot sensitivity curves for direct detection experiments in the
$(m_{\tilde{\chi}}, \sigmaSI^p )$ plane, along with predictions for the models we discuss with $m_{\tilde s_1} = 2~\tev$, $m_{\tilde s_2}=10$ TeV, and $\sin (2\phi_{\tilde{s}})=1$.
Again, the black line is the bound from
current LUX data~\cite{LUX}, while the dashed red line is the sensitivity curve for LUX with 300 days of data, and the
blue dot-dashed curve could be probed by LZ-7~\cite{LUXproj}.
The green bands indicate the uncertainty in the scattering cross section for the above model as a function of the strangeness content
of the nucleon.  The dark green band indicates the predicted SI-scattering cross section
if one takes $\sigma_0 =27~\mev$ and allows the full $2\sigma$ range for $\Sigma_{\pi N}$ of 45 MeV to 73 MeV~\cite{Alarcon:2011zs}
The light green band indicates the predicted spin-independent scattering cross section if one decreases the strangeness content
further, to a minimum of $B_s^N =0.5$ (which, again, is still larger than the value favored by lattice calculations and the value used by default in micrOMEGAS versions 3.0 through 3.5.5~\cite{Belanger:2013oya}).
Note that the
spin-independent scattering cross section scales as $\sin^2 (2\phi_{\tilde{s}})$; for a different choice one can simply rescale
the cross section appropriately.  Similarly, in the $m_{\tilde{\chi}} \ll m_{\tilde s_1}$ limit, the cross section
scales as $m_{\tilde s_1}^{-4}$.  Again we see that the sensitivity of direct detection experiments grows
rapidly as one approaches the limit $m_{\tilde{\chi}} \sim m_{\tilde s_1}$, but this effect will be cut off by the
momentum transfer of the scattering process, as discussed above.

We have demonstrated that if spin-independent scattering is mediated by strange squarks, then the sensitivity of direct detection experiments
depends crucially on the strangeness content of the nucleon, and can result in scattering cross sections that vary over many orders of magnitude, a conclusion which is further strengthened if one considers the even smaller nucleon form factors favored by lattice calculations.  We now turn briefly to the case of scattering mediated by up or down squarks.  An important difference between scattering mediated by strange squarks vs.~up or down squarks is that in the former case the scattering is isospin invariant, while the latter scatterings are isospin-violating.  If the only non-decoupled squarks are first generation squarks  then the sensitivities
of direct detection experiments would have to be rescaled to account for the resulting isospin violation.  That said, unless isospin violation yields significant destructive interference, the resulting degradation in
sensitivity is typically an ${\cal O}(1)$ factor.

As mentioned in Sec.~\ref{sec:scatteringeqs}, approximate cross sections for the cases of light up or down squarks can be obtained by rescaling Equations~\ref{eq:SigmaSI_approx} or~\ref{eq:SigmaSI_resonant} with the appropriate hypercharge, nucleon form factor, and squark mixing angle.
We see that next generation direct
detection experiments can probe models that are consistent with LHC data and that have $\sin (2\phi_{\tilde{u},\tilde{d}}) \ll 1$.
Referring to the $2\sigma$ ranges for $\Sigma_{\pi N}$ in Table~\ref{tab:Buds}, we see that even if first generation squarks mediate bino-nucleon scattering, there is a factor of $\sim2$ uncertainty in $B_{u,d}^N$, leading to a factor of $\sim4$ uncertainty in $\sigma_{\rm SI}^N$.  In this case, the uncertainty comes mainly from the value of $\Sigma_{\pi N}$, though the strangeness content of the nucleon still plays a significant role.

\section{Conclusions}

We have considered the sensitivity of direct detection experiments to the spin-independent
scattering of bino-like dark matter in a scenario within the MSSM in which some assumptions that
lead to CMSSM-based intuition are relaxed.  Our analysis has been undertaken from a low energy point of view, allowing $L$-$R$ squark (or slepton) mixing, and allowing the squark masses to take any values currently allowed by experimental constraints.  In this scenario we have identified three decoupled
sectors of the parameter-space which control different aspects of the physics: the leptonic
sector controls the thermal relic density, the heavy squark sector controls the Higgs boson mass,
and the light squark sector controls dark matter-nucleon scattering.
For low-mass dark matter, this scenario thus corresponds to an extension of the ordinary CMSSM ``bulk" region of parameter space,
as discussed in~\cite{Fukushima:2014yia}, where bino annihilation via slepton exchange is responsible
for sufficiently depleting the dark matter relic density.  For heavier dark matter, this scenario instead corresponds to a co-annihilation
region, where the dark matter relic density is depleted through dark matter co-annihilation with a light squark.
We have shown here that for
models with significant left-right light squark mixing, and which are consistent with all other
data, spin-independent dark matter-nucleon scattering can arise from squark exchange and can
be observed with direct detection experiments such as LUX.

The connection between model parameters and observed rates at direct detection experiments are
sensitive to nuclear physics uncertainties, and in particular to the quark content of the
nucleon.  Recent results from lattice QCD suggest that the strange quark content of the nucleon
is negligible, in which case light up or down squarks will likely be the dominant mediators of spin-independent scattering interactions.
If the strange quark content is non-trivial, however, then light strange squark exchange can
dominate spin-independent scattering.

Spin-independent scattering in this scenario defies some of the common lore associated with the more typical MSSM (or CMSSM) case, in which the dominant process is Higgs exchange.   For example, in this scenario, spin-independent scattering need not be isospin-invariant,
and in general $f_n / f_p$ can assume any value, including negative values.  This opens the
possibility, within the context of the MSSM, for destructive interference which significantly
reduces the sensitivity of xenon- or germanium-based direct detection experiments, thus highlighting
the necessity of multiple detector target materials.  It is interesting to note that if data from
the LHC should suggest that either up or down squarks are light, while all others are heavy, then
this scenario would provide a prediction for $f_n / f_p = B_{u,d}^n / B_{u,d}^p$.
Though the magnitude of either $f_n$ or $f_p$ would depend on the light squark left-right mixing angle,
the ratio does not.  If indeed
$B_s^N \sim 0$, as suggested by lattice QCD results, then the prediction for $f_n / f_p$ for
up or down squark exchange is $1/z$ or $z$, respectively, where $z = 1.49$ is determined precisely
from baryon mass measurements.  This testable prediction would thus present a nice correlation of
data from the LHC, direct detection experiments, and meson spectroscopy.

It is also interesting to note that the bino-nucleon scattering cross section exhibits an enhancement
when $m_{\tilde{\chi}} \sim m_{\tilde f_1}$.  This is also the region of parameter space in which bino
coannihilation with light squarks can enhance the depletion of dark matter in the early universe,
allowing the bino thermal relic density to be consistent with astronomical observations.
The ordinary ``coannihilation" region of CMSSM parameter space involves neutralino coannihilation
with stops or staus; thus, we see that the extension of the CMSSM bulk region, which we focus on,
naturally connects to an extension of the CMSSM coannihilation region (this time involving $\tilde u$, $\tilde d$,
or $\tilde s$).

Next generation direct detection experiments, such as LZ-7, may be sensitive to spin-independent scattering via
squark exchange even if the squarks are as heavy as several TeV.  This sensitivity rises even more dramatically
if the dark matter and squark are nearly degenerate in mass.
For such heavy squarks, with $m_{\tilde{q}} \gtrsim 3$ TeV, direct production
at the LHC would be very difficult, and even indirect signatures may be hard to identify.  In such cases, the first signs of squarks may arise from direct dark matter detection experiments.
Although we have analyzed a MSSM scenario, this lesson is more general.  The LHC excels at producing QCD-charged
particles, and if new heavy QCD-charged particles exist then it is usually assumed that the LHC is the experiment
best suited for discovering them.  However, LHC sensitivity is sharply curtailed above the kinematic threshold for
particle pair production.  If very heavy QCD-charged particles couple to dark matter, however, low-background
dark matter direct detection experiments may provide a complementary tool to high energy collider experiments.

Finally, we would like to draw attention to the detailed correlation between the dark matter-nucleon scattering cross section
and the quark content of the nucleon.  For example, if it is determined that bino-nucleon spin-independent scattering is dominated by
the exchange of one flavor of first generation squarks, then the measurement of the relative strength of bino-proton and bino-neutron
interactions determines the ratios $B_s^{N} / B_{u,d}^{N}$.  If dark matter couples to nucleons via the exchange of non-degenerate
squarks, then it provides a probe of the structure of the nucleon that is completely different from any other probe provided
by the Standard Model, potentially allowing the possibility of using dark matter to study nuclear physics.

In the meantime, we encourage both investigations of the quark content of the nucleon and attention to these fundamental uncertainties that underly interpretations of results from direct dark matter searches.

\vskip .2in
%\newpage
{\bf Acknowledgments.}
We are grateful to Jonathan Feng, Azar Mustafayev and Xerxes Tata for useful discussions.  We are especially grateful to
Nicolas Fernandez for collaboration at an early stage of this project.
The work of P.~Stengel is supported in part by DOE grant DE-SC0010504. The work of J.~Kumar is supported in part by NSF CAREER Award No. PHY-1250573.
The work of P.~Sandick is supported in part by NSF Grant No.~PHY-1417367.
J.~Kumar and P.~Sandick  would like to thank the University of Utah and University of Hawaii, respectively, for their hospitality and partial support.
C.~Kelso, J.~Kumar, and P.~Sandick would also like to thank CETUP* (Center for Theoretical Underground Physics and Related Areas),
supported by the US Department of Energy under Grant No. DE-SC0010137 and by the US National Science Foundation
under Grant No.~PHY-1342611, for its hospitality and partial support, where first
discussions of this idea took place.

\appendix
\section{MSSM Mass Matrices}

We can restate the expressions for squark-mediated neutralino scattering in a supersymmetric framework within the MSSM, following~\cite{Falk:1998xj}. In general, the neutralino LSP is a mass eigenstate composed of the bino $\tilde{B}$, wino $\tilde{W}$, and higgsinos $\tilde{H}_{1,2}$ and the mass matrix can be written in the $( \tilde{B}, \tilde{W}^3, \tilde{H}_1^0 , \tilde{H}_2^0 ) $ basis as
\bea
\begin{bmatrix}
M_1 &  0 & - M_Z \sin \theta_W \cos \beta & M_Z \sin \theta_W \sin \beta \\
0 &  M_2 & M_Z \cos \theta_W \cos \beta & - M_Z \cos \theta_W \sin \beta \\
- M_Z \sin \theta_W \cos \beta &  M_Z \cos \theta_W \cos \beta  & 0 & - \mu \\
M_Z \sin \theta_W \sin \beta & - M_Z \cos \theta_W \sin \beta & - \mu & 0
\end{bmatrix}.
\eea
The squared sfermion mass matrix can be written as
\bea
\small
M^2 =
\begin{bmatrix}
M_L^2 + m_f^2 + \cos 2 \beta \left( T_{3f} - Q_f \sin^2 \theta_W \right) M_Z^2 &  - m_f  \bar{m}_f  e^{\imath \gamma_f}   \\
 m_f  \bar{m}_f  e^{- \imath \gamma_f} & M_R^2 + m_f^2 + \cos 2 \beta  Q_f \cos^2 \theta_W  M_Z^2
\end{bmatrix}
\label{eq:SfermionMassMatrix},
\eea
where $M_{L (R)}$ are the soft SUSY breaking masses, assumed to be diagonal. As we allow for left-right sfermion mixing, the off diagonal terms are defined
\bea
\bar{m}_f e^{\imath \gamma_f} &=& \mu R_f + A_f^* = R_f | \mu | e^{\imath \theta_\mu} + | A_f | e^{ - \imath \theta	_{A_f}},
\eea
where $m_f$ is the fermion mass and $R_f = \cot \beta$ for weak isospin $1/2$ fermions ($R_f = \tan \beta$ for weak isospin $- 1/2$). We diagonalize the squared sfermion  mass matrix with ${\rm diag}( m_1^2 , m_2^2 ) = \eta M^2 \eta^{-1}$,  with
\bea
\eta =
\begin{bmatrix}
\cos \theta_f &  \sin \theta_f e^{\imath \gamma_f} \\
- \sin \theta_f e^{- \imath \gamma_f} &  \cos \theta_f
\end{bmatrix}
=
\begin{bmatrix}
\eta_{1 1}  &  \eta_{1 2} \\
\eta_{2 1} &  \eta_{2 2}
\end{bmatrix},
\eea
parametrized by the mixing angle $\theta_f$ for each squark flavor, in the notation of~\cite{Ellis:2008hf}. For a pure bino LSP, the effective coupling for the scalar mediated  4-point interaction is given by
\bea
\alpha_i &=&
 - {  Re \left[ \left( X_i \right) \left( Y_i \right)^* \right] \over 2 \left( m_{\tilde{q}_{1 i}} - m_{\tilde{\chi}}^2 \right)}
 - {  Re \left[ \left( W_i \right) \left( V_i \right)^* \right] \over 2 \left( m_{\tilde{q}_{2 i}} - m_{\tilde{\chi}}^2 \right)},
\eea
with
\bea
X_i &=& - \eta_{12}^* e_i g '
\nonumber \\
Y_i &=& \eta_{11}^* {y_i \over 2} g '
\nonumber \\
W_i &=& - \eta_{22}^* e_i g '
\nonumber \\
V_i &=& \eta_{21}^* {y_i \over 2} g ',
\eea
where $i$ labels up-type ($i=1$) or down-type ($i =2$) quarks. Note that $y_i$, $e_{i}$ and $g '$ are the hypercharge, electromagnetic charge and hypercharge coupling, respectively.

\end{document}